\documentclass[twocolumn]{revtex4} % for arxiv
\usepackage{epsfig}
\usepackage{graphicx}
\usepackage{dcolumn}
\usepackage{bm}
\usepackage{braket}
\usepackage{amssymb}
\usepackage{amsmath}

\begin{document}

\title{Impedance and Scattering Variance Ratios of Complicated Wave Scattering Systems in the Low Loss Regime}

\author{Jen-Hao Yeh$^{a}$}
%\email[]{davidyeh@umd.edu}
\author{Zachary Drikas$^{b}$}
\author{Jesus Gil Gil$^{b}$}
\author{Sun Hong$^{b}$}
\author{Biniyam T. Taddese$^{a}$}
\author{Edward Ott$^{a}$}
\author{Thomas M. Antonsen$^{a}$}
\author{Tim Andreadis$^{b}$}
\author{Steven M. Anlage$^{a}$}

\affiliation{$^{a}$University of Maryland, College Park, MD 20742,
US} \affiliation{$^{b}$U.S. Naval Research Laboratory, Washington,
DC 20375, US}

\begin{abstract}
Random matrix theory (RMT) successfully predicts universal
statistical properties of complicated wave scattering systems in the
semiclassical limit, while the random coupling model offers a
complete statistical model with a simple additive formula in terms
of impedance to combine the predictions of RMT and nonuniversal
system-specific features. The statistics of measured wave properties
generally have nonuniversal features. However, ratios of the
variances of elements of the impedance matrix are predicted to be
independent of such nonuniversal features and thus should be
universal functions of the overall system loss. In contrast with
impedance variance ratios, scattering variance ratios depends on
nonuniversal features unless the system is in the high loss regime.
In this paper, we present numerical tests of the predicted universal
impedance variance ratios and show that an insufficient sample size
can lead to apparent deviation from the theory, particularly in the
low loss regime. Experimental tests are carried out in three
two-port microwave cavities with varied loss parameters, including a
novel experimental system with a superconducting microwave billiard,
to test the variance-ratio predictions in the low loss
time-reversal-invariant regime. It is found that the experimental
results agree with the theoretical predictions to the extent
permitted by the finite sample size.
\\


\end{abstract}

\maketitle

\section{Introduction}
% Motivation
Understanding the properties of complicated wave scattering systems
\cite{Newton1966B} is a common challenge in many engineering and
physics fields, such as quantum chaotic systems
\cite{Stockmann1999B, Haake2000B}, quantum dots and mesoscopic
systems \cite{Altshuler1991B, Brouwer1997, Alhassid2000,
Mello2004B}, acoustic waves \cite{Pagneux2001}, and microwave
cavities \cite{Doron1990, Kuhl2005, Hemmady2005a, Hemmady2005b}. Due
to the complexity of wave propagation and scattering in many of
these systems, numerically solving the wave equations with high
resolution is difficult or impractical. This is particularly true
when the wavelength is short compared to the characteristic size of
the scattering region (the situation of interest in this paper). In
addition, in this case, scattering properties are extremely
sensitive to small changes in system parameters, which may not be
precisely known. Thus, a statistical approach has become a popular
alternative for describing the wave properties \cite{Holland1999B}.

% Random matrix theory and problem
Researchers have developed statistical models based on random matrix
theory (RMT), which successfully predict certain universal
statistical properties of complicated wave scattering systems
\cite{Bohigas1984, Mehta1991B, Akemann2011B}. In order to apply RMT
to practical wave systems, one usually needs to account for
nonuniversal system-specific features, which are not included in
RMT. For example, considering microwave signals entering an
enclosure through localized ports and propagating inside, the port
coupling between the enclosure and the outside world is one
system-specific feature \cite{Zheng2006a,Zheng2006b}. The short ray
trajectories between ports due to scattering from fixed walls and/or
objects within the enclosure are also nonuniversal system-specific
features \cite{Hart2009}.

% RCM and impedance variance ratio
The random coupling model (RCM) is a well-developed model to combine
the universal predictions of RMT and the nonuniversal features of a
practical system by a simple additive formula in terms of impedance
\cite{Zheng2006a,Zheng2006b,Hart2009}. This model has been
experimentally verified in microwave cavities, and it offers a
complete statistical model for the impedance matrices, the
scattering matrices \cite{Hemmady2005a, Hemmady2005b, Yeh2010a,
Yeh2010b}, the admittance matrices \cite{Hemmady2006a}, the
conductances \cite{Hemmady2006c}, and the fading statistics
\cite{Yeh2012a,Yeh2012b} of practical systems. The statistical
distributions of the universal predictions of RMT and the practical
distributions which includes nonuniversal features are distinctly
different for most wave scattering properties. However, the
impedance variance ratio (defined below) is a quantity that is
predicted to be independent of nonuniversal features of the wave
system, and it is expected to be a universal function of the loss of
the system \cite{Zheng2006c}. In this paper, we use ``universality'' 
to mean that the impedance variance ratio is independent of the 
system-specific features including the port coupling of the system 
and the short ray trajectories betweeen ports.

% Z and S
Impedance is a meaningful concept in electromagnetism, and it can be
extended to all wave scattering systems. In a linear electromagnetic
wave system with $N$ ports, the $N\times N$ impedance matrix
$\textbf{Z}$ is the linear relationship of the complex phasor
voltage vector $\widehat{\textbf{\textit{V}}}$ of the $N$ port
voltages and the complex phasor current vector
$\widehat{\textbf{\textit{I}}}$ of the $N$ port currents, via the
phasor generalization of Ohm's law as $
\widehat{\textit{\textbf{V}}}=\textbf{Z}\widehat{\textit{\textbf{I}}}$
\cite{Pozar1990B}. A quantum-mechanical quantity corresponding to
the impedance is the so-called reaction matrix, which is often
denoted in the literature as $\textbf{K}$ and is related to
$\textbf{Z}$ by $\textbf{K}=-i\textbf{Z}$ \cite{Alhassid2000,
Verbaarschot1985, Lewenkopf1991, Beck2003, Fyodorov2004,
Fyodorov2005, Savin2005}. The impedance matrix can also be related
to the scattering matrix $\textbf{S}$ via the relationship
\cite{Zheng2006a,Zheng2006b}
\begin{equation}\label{eq:StoZ}
    \textbf{Z} = \textbf{Z}_{0}^{1/2}\left(\textbf{1}+\textbf{S}\right)\left(\textbf{1}-\textbf{S}\right)^{-1}\textbf{Z}_{0}^{1/2},
\end{equation}
where $\textbf{Z}_{0}$ is a $N\times N$ diagonal matrix whose
diagonal element $Z_{0,nn}$ is the characteristic impedance of the
n$^{th}$ scattering channel mode, and $\textbf{1}$ is the identity
matrix. The scattering matrix $\textbf{S}$ specifies the linear
relationship between the incoming wave vector
$\widehat{\textit{\textbf{a}}}$ and the outgoing wave vector
$\widehat{\textit{\textbf{b}}}$, as
$\widehat{\textit{\textbf{b}}}=\textbf{S}\widehat{\textit{\textbf{a}}}$.
The $n^{th}$ element of the incoming and outgoing power waves are
$a_{n}=(V_{n}+Z_{0,nn}I_{n})/\sqrt{Z_{0,nn}}$ and
$b_{n}=(V_{n}-Z_{0,nn}I_{n})/\sqrt{Z_{0,nn}}$, where $V_{n}$ and
$I_{n}$ are the voltage and current at the $n^{th}$ port,
respectively \cite{Pozar1990B}, and the incident and reflected power
fluxes in channel $n$ are $|a_{n}|^{2}$ and $|b_{n}|^{2}$.

% variance statistics
For complicated wave scattering systems, the impedance matrices and
the scattering matrices are sensitive to small variations of the
system, such as change of the applied frequency, the configuration
of the enclosure boundary, or the location and orientation of an
internal scatterer. The statistical variations of the elements of
$\textbf{Z}$ and $\textbf{S}$ due to small random changes in the
scattering system are of great interest
\cite{Zheng2006c,Fyodorov2005,Savin2006}. For example, the variances
of the the elements of $\textbf{S}$ and their ratio (the
Hauser-Feshbach relation) have been studied in the nuclear
scattering literature when researchers investigate the statistics of
inelastic scattering of neutrons \cite{Hauser1952} and compound
nuclear reactions \cite{Verbaarschot1985,Agassi1975}. Friedman and
Mello used information theory to derive the Hauser-Feshbach formula
in the statistical treatment of nuclear reactions
\cite{Friedman1985}.

The elastic enhancement factor is the ratio of variances in
reflection (diagonal elements of $\textbf{S}$) to that in
transmission (off-diagonal elements of $\textbf{S}$)
\cite{Verbaarschot1986}. In chaotic scattering, elastic processes
(the diagonal elements) are known to be systematically enhanced over
inelastic ones (the off-diagonal elements) \cite{Kretschmer1978,
Dietz2010}. For a two-port system, the elastic enhancement factor
$W=\sqrt{\textrm{Var}[S_{11}]\textrm{Var}[S_{22}]}/\textrm{Var}[S_{12}]$,
where $\textrm{Var}[x]$ stands for the variance of the variable $x$,
and $S_{ij}$ denotes the matrix element of $\textbf{S}$ that
occupies the $i^{th}$ row and the $j^{th}$ column. In research on
electromagnetic fields in mode-stirred reverberating chambers,
Fiachetti and Michelsen have conjectured the universality of the
ratio of the variances of the scattering elements in the cases of
time reversal invariant systems (corresponding to RMT of the
Gaussian orthogonal ensemble (GOE)) \cite{Fiachetti2003}. The
universality of the scattering variance ratio has been tested with
wave scattering experiments in microwave resonators in the GOE case
\cite{Zheng2006c}. Dietz \textit{et al.} have also tested the
universality of the elastic enhancement factor with microwave
resonators in the GOE case and in the cases of partially breaking of
time reversal invariance (corresponding to RMT of the Gaussian
unitary ensemble (GUE)) \cite{Dietz2010}. {\L}awniczak \textit{et
al.} have used microwave networks to test the elastic enhancement
factor in both the GOE and GUE cases
\cite{Lawniczak2010,Lawniczak2011,Lawniczak2012}.

% impedance variance ratio
In this paper we are concerned with the impedance variance ratio,
which is defined as \cite{Zheng2006c}
\begin{equation}\label{eq:variance_ratioZ}
    \Xi_{Z}\equiv\frac{\textrm{Var}[Z_{ij}]}{\sqrt{\textrm{Var}[Z_{ii}]\textrm{Var}[Z_{jj}]}},\
\ \  i\neq j,
\end{equation}
and the scattering variance ratio, defined as
\begin{equation}\label{eq:variance_ratioS}
    \Xi_{S}\equiv\frac{\textrm{Var}[S_{ij}]}{\sqrt{\textrm{Var}[S_{ii}]\textrm{Var}[S_{jj}]}},\
\ \  i\neq j,
\end{equation}
where the variances arise from small variations of the system. For a
reciprocal ($Z_{ij}=Z_{ji}$) two-port system, the impedance variance
ratio is
$\Xi_{Z}=\textrm{Var}[Z_{12}]/\sqrt{\textrm{Var}[Z_{11}]\textrm{Var}[Z_{22}]}$.
Similarly, the scattering variance ratio is
$\Xi_{S}=\textrm{Var}[S_{12}]/\sqrt{\textrm{Var}[S_{11}]\textrm{Var}[S_{22}]}$.
Note that $\Xi_S$ is the inverse of the elastic enhancement factor
of a two-port system. The impedance variance ratio $\Xi_{Z}$ is
predicted to be a universal function of the loss parameter $\alpha$
\cite{Zheng2006c}, which characterizes the losses and mode-spacing
within the wave scattering system (defined below). On the other
hand, $\Xi_{S}$ is in general dependent on the system-specific
features of the wave scattering system and hence not universal. Only
in the high loss regime ($\alpha \gg 1$) can one assume that the
fluctuating part of the impedance matrix (or the scattering matrix)
is much smaller than the mean part, which allows one can obtain the
result $\Xi_{S}\simeq\Xi_{Z}$ ($\alpha\gg 1$) \cite{Zheng2006c},
which implies that $\Xi_{S}$ is approximately universal for high
loss.

% loss parameter
The loss parameter can be understood as the degree of overlap of
resonances in frequency in the electromagnetic case (or energy level
in the quantum case) due to the distributed losses of the closed
version of the wave scattering system. For example, in the case of
electromagnetic wave scattering, the loss parameter is
\begin{equation}\label{eq:loss_para}
    \alpha=\frac{f}{2Q\Delta f},
\end{equation}
where $f$ is the frequency of the wave signal, $\Delta f$ is the
average spacing between cavity resonant frequencies near $f$, and
$Q$ is the quality factor due to the distributed losses of the
closed cavity, such as losses from conducting walls or a lossy
dielectric that fills the cavity \cite{Hemmady2005a,
Zheng2006a,Zheng2006b}. Based on RMT, researchers have given 
analytical expressions of $\Xi_{Z}(\alpha)$ \cite{Zheng2006c,Fyodorov2005} 
and $\Xi_{S}(\alpha)$ \cite{Fyodorov2005,Savin2006} for the GOE and GUE
cases. In this paper, we focus on the
time reversal invariant case (GOE).

% Our gaols
The goal of this paper is to experimentally test the analytical
predictions of the impedance variance ratio and the scattering
variance ratio in the low loss regime. Dietz \textit{et al.} carried
out experiments in the low loss regime, but their interests were in
the elastic enhancement factor (inverse of $\Xi_{S}$) in the weak
port-coupling situation \cite{Dietz2010}. Note that the common
approach to accounting for coupling (one nonuniversal feature) is to
use a single scalar quantity for a given frequency range (the
amplitude of the averaged scattering parameter
$|\overline{S_{ii}}|$) \cite{Kuhl2005, Savin2006}, whereas the
random coupling model treats nonuniversal features more generally by
using a complex function of frequency (the frequency-dependent
averaged impedance matrix, defined in Section 2.2), and includes
short ray trajectories. Zheng \textit{et al.}'s study of $\Xi_{Z}$
and $\Xi_{S}$ \cite{Zheng2006c} and {\L}awniczak \textit{et al.}'s
study of the elastic enhancement factor
\cite{Lawniczak2010,Lawniczak2011,Lawniczak2012} applied the
original version of the RCM to take account of the nonuniversality
of the port-coupling. In this paper we apply the extended version of
the RCM to further include the nonuniversal features of the short
ray trajectories. We also test the low loss regime which has not
been previously achieved by Zheng's or {\L}awniczak's experiments
\cite{Zheng2006c, Lawniczak2010,Lawniczak2011,Lawniczak2012}.

In the following sections, we first review the theory and present
numerical tests of $\Xi_{Z}$ and $\Xi_{S}$ as a function of loss
parameter. The numerical tests point out a numerical deviation from
the theory due to the finite number of samples, which is more
significant in the low loss regime for the impedance variance ratio.
After the numerical tests, we present our experimental systems of
three microwave cavities with varied values of the loss parameter
and make a thorough experimental test in a broad range of loss
parameters.

\section{Theory and Numerical Results}

\subsection{Universal Statistics Based on RMT}

% RMT impedance
The theoretical model of the impedance variance ratio $\Xi_{Z}$ is
derived from RMT \cite{Zheng2006c}. Using RMT, for a complicated
wave scattering system with time reversal invariance of wave
propagation, researchers have developed a statistical model of the
impedance matrix $\textbf{Z}_{rmt}$ \cite{Zheng2006a, Zheng2006b,
Hart2009, Lewenkopf1991, Beck2003, Fyodorov2004, Fyodorov2005,
Savin2005}. This statistical model is applicable to situations where
system-specific short-ray-trajectory effects are negligible and the
ports are such that the input-output channels are perfectly matched
to the scatterer (in the sense that $\langle \textbf{Z} \rangle =
\textbf{1}$, where $\langle\ldots \rangle$ denotes a suitable
ensemble average).

% RMT VRZ
With the known statistics of $\textbf{Z}_{rmt}$, the impedance
variance ratio as a function of $\alpha$ can be analytically derived
\cite{Mehta1991B,Zheng2006a}
\begin{equation}\label{eq:VRZ_analy}
    \Xi_{Z_{rmt}}(\alpha)=\left[3-2\int_{0}^{\infty}\frac{4\ g(x)}{4+(x/\alpha)^{2}}\ dx
    \right]^{-1},
\end{equation}
where $\displaystyle g(x) =
f^{2}(x)-\left[\int^{x}_{0}f(x')dx'-\frac{1}{2}\right]\frac{df}{dx}$
and $\displaystyle f(x)=\frac{\sin(\pi x)}{\pi x}$ in the time
reversal invariant case. This result is shown as the thick black
curve in Fig.$\ $1, where the loss parameter scale is logarithmic.
Note that $\Xi_{Z_{rmt}} = 1/3$ in the GOE lossless case ($\alpha =
0$) and $\Xi_{Z_{rmt}} = 1/2$ as $\alpha \rightarrow \infty$.

\begin{figure}
\includegraphics[width=3.2in]{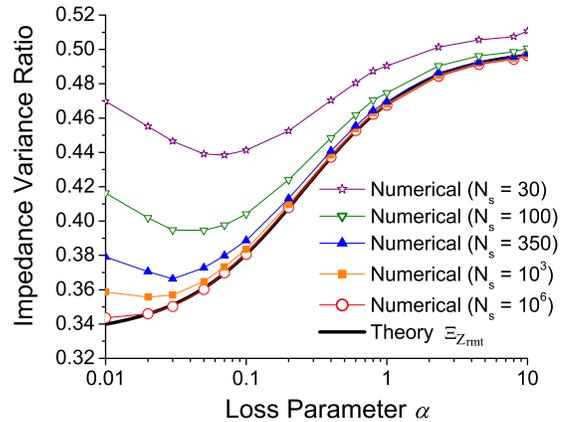}

\caption{The impedance variance ratio versus the loss parameter
$\alpha$. The thick black curve is the analytical formula
$\Xi_{Z_{rmt}}$, Eq.$\ $(5). The other colored curves are numerical
results of mean impedance variance ratio
$\widetilde{\Xi}_{Z_{rmt}}^{(N_{s})}$ based on $\textbf{Z}_{rmt}$
with different numbers of samples ($N_{s}$) indicated in the
parentheses.}

\end{figure}

% numerical VRZ
In addition to the analytical prediction (Eq.$\ $(5)), we also
numerically generate $2\times 2$ random impedance matrices
$\textbf{Z}_{rmt}$ (using the appropriate RMT ensemble) and compute
the variance ratios with different values of the loss parameter
$\alpha$. We select 15 different loss parameters from $\alpha =
0.01$ to $\alpha = 10$. With each loss parameter, we generate a
finite ensemble with $N_{s}$ samples of $\textbf{Z}_{rmt}$ matrices.
The variations of these matrices represent a finite sampling of the
universal variations of the wave scattering system. Because the
number of generated sample matrices ($N_{s}$) is finite, the
variance ratio $\displaystyle\Xi_{Z_{rmt}}^{(N_{s})}=
\frac{\textrm{Var}^{(N_{s})}[Z_{rmt,12}]}{\sqrt{\textrm{Var}^{(N_{s})}[Z_{rmt,11}]\textrm{Var}^{(N_{s})}[Z_{rmt,22}]}}$
of a finite ensemble is not a single value, but has a statistical
distribution. To illustrate the finite-sample-size issue, we choose
the sample numbers as $N_{s}=$ 30, 100, 350, $10^{3}$, and $10^{6}$
for each loss parameter, and we numerically generate the statistical
distribution of $\Xi_{Z_{rmt}}^{(N_{s})}$. We plot the means of
these distributions ($\widetilde{\Xi}_{Z_{rmt}}^{(N_{s})} =
\langle\Xi_{Z_{rmt}}^{(N_{s})}\rangle$) versus the loss parameter as
colored curves in Fig.$\ $1. One can see the deviations between the
numerical $\widetilde{\Xi}_{Z_{rmt}}^{(N_{s})}$ and the analytical
theory (Eq.$\ $(5)) are more significant in the low loss cases. This
indicates that fluctuations of $\Xi_{Z_{rmt}}^{(N_{s})}$ in the low
loss cases are more significant, thus necessitating a large number
of samples to achieve good agreement between the finite-size
numerical mean and the theory.

% numerical VRS
As with the impedance variance ratio $\Xi_{Z_{rmt}}$, we have done
the same analysis for the scattering variance ratio $\Xi_{S_{rmt}}$,
where
$\textbf{S}_{rmt}=(\textbf{Z}_{rmt}-\textbf{1})(\textbf{Z}_{rmt}+\textbf{1})^{-1}$.
For the scattering matrices generated based on RMT in the time
reversal invariant (GOE) case, the theoretical prediction is
$\Xi_{S_{rmt}} = 1/2$ \cite{Brouwer1997,Zheng2006c}, and it is
independent of the loss parameter $\alpha$. We show the theory and
numerical results in Fig.$\ $2. Note that $\Xi_{S_{rmt}}$ does not
contain the nonuniversal features encountered in a typical practical
system.

\begin{figure}

\includegraphics[width=3.2in]{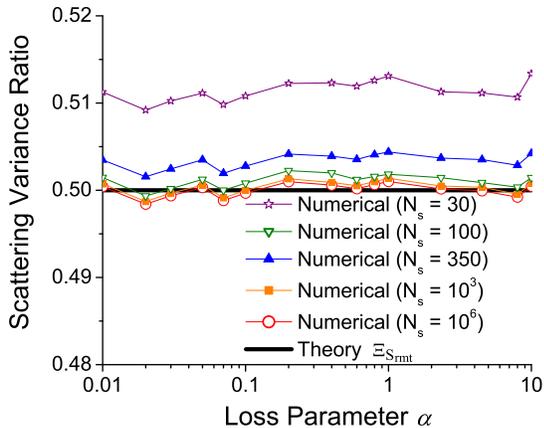}

\caption{The scattering variance ratio versus the loss parameter
$\alpha$. The thick black curve is the theory $\Xi_{S_{rmt}} = 1/2$.
The other colored curves are numerical results
$\widetilde{\Xi}_{S_{rmt}}^{(N_{s})}$ with different numbers of
samples ($N_{s}$) indicated in the parentheses.}

\end{figure}

\subsection{Including the Nonuniversal Features through the RCM}
% RCM
To extend the predictions of RMT to practical systems and include
nonuniversal features, Zheng \textit{et al.} have introduced the
random coupling model \cite{Zheng2006a,Zheng2006b}. The original
version of the random coupling model took the system-specific port
coupling into account through the radiation impedance matrix. This
method has also been applied in previous work on impedance and
scattering variance ratios \cite{Zheng2006c}. Hart \textit{et al.}
have considered the additional system-specific features of short ray
trajectories between ports and developed the
short-ray-trajectory-corrected version of the RCM \cite{Hart2009}.
This RCM connects the universal fluctuating part and the practical
impedance matrix $\textbf{Z}$ as
\begin{equation}\label{eq:RCMavg}
\textbf{Z}_{n}=\textbf{R}_{avg}^{-1/2}
\left(\textbf{Z}-i\textbf{X}_{avg}\right) \textbf{R}_{avg}^{-1/2}.
\end{equation}
The normalized impedance matrix $\textbf{Z}_{n}$ represents the
universal part, and its statistics are the same as the RMT
prediction ($\textbf{Z}_{rmt}$) \cite{Yeh2010a, Yeh2010b}. The
nonuniversal features of the port coupling (the radiation impedance)
and short ray trajectories are included in the ensemble-averaged
impedance matrix
$\textbf{Z}_{avg}=\textbf{R}_{avg}+i\textbf{X}_{avg}$, where
$\textbf{R}_{avg} = \textrm{Re}[\textbf{Z}_{avg}]$,
$\textbf{X}_{avg} = \textrm{Im}[\textbf{Z}_{avg}]$ \cite{Hart2009,
Yeh2010b}.

% ensemble
In experiments measuring the statistics of wave scattering
properties, one needs an ensemble measurement of many different
realizations of the system \cite{Hemmady2005a, Yeh2010a, Yeh2010b,
Schafer2005, Schanze2005}. In this paper, our experimental
measurement ensemble includes configuration variation and frequency
variation. These variations aim to create a set of systems in which
none of the nonuniversal system details are reproduced from one
realization to another, except for the effects of the port coupling
and short ray trajectories. The previous analysis of the
experimental results for the impedance variance ratio included
frequency-dependent nonuniversal feature of short ray trajectories
\cite{Zheng2006c}. In this paper we remove these by utilizing the
extended RCM (Eq.$\ $(6)) \cite{Yeh2010a, Yeh2010b}.

% assumption for \Xi_{Z}
Considering the extended RCM (Eq.$\ $(6)), in general the variance
ratio $\Xi_{Z}$ of the impedance matrix and the variance ratio
$\Xi_{Z_{n}}$ of the normalized impedance matrix are not equal, and
their relationship depends on the elements of $\textbf{R}_{avg}$
(note that $\textbf{X}_{avg}$ does not influence the variances of
the impedance elements). However, if the ports of the wave
scattering system are far apart, then the off-diagonal elements of
$\textbf{Z}_{avg}$ are small \cite{Zheng2006c}, and one can
approximately simplify the relationship between $\Xi_{Z}$ and
$\Xi_{Z_{n}}$. More specifically for a two-port system, one can
define
\begin{equation}\label{eq:Rsqrt}
\textbf{R}_{avg}^{1/2} = \left[
       \begin{array}{cc}
         A & B \\
         C & D \\
       \end{array}
     \right],
\end{equation}
where $A$, $B$, $C$, and $D$ ($B=C$ in time reversible (reciprocal)
cases) are all frequency-dependent real quantities. Under the
condition $A$, $D\gg |B|$, $|C|$, the relationships of impedance
variances over configuration realizations at a frequency $f$ become
\begin{equation}\label{eq:variance_11}
\textrm{Var}[Z_{11}]=A^{2}(f)\textrm{Var}[Z_{n,11}],
\end{equation}
\begin{equation}\label{eq:variance_22}
\textrm{Var}[Z_{22}]=D^{2}(f)\textrm{Var}[Z_{n,22}],
\end{equation}
\begin{equation}\label{eq:variance_12}
\textrm{Var}[Z_{12}]=A(f)D(f)\textrm{Var}[Z_{n,12}].
\end{equation}
In this case, $A(f)$ and $D(f)$ cancel in the calculation of the
variance ratio, and one has the universal result
\begin{equation}\label{eq:variance_equal}
\Xi_{Z} = \Xi_{Z_{n}}.
\end{equation}
This equation shows the significance of the impedance variance
ratio: if off-diagonal elements of $\textbf{R}_{avg}$ are
negligible, the quantity is independent of the system-specific
feature $\textbf{Z}_{avg}$ and is directly related to the universal
fluctuating quantity $\textbf{Z}_{n}$. The statistics of
$\textbf{Z}_{n}$ are the same as the statistics of
$\textbf{Z}_{rmt}$, and the statistical properties only depend on
the loss parameter $\alpha$ \cite{Zheng2006c}. Therefore, the
impedance variance ratio becomes is a universal property of the wave
scattering system and only depends on the loss parameter $\alpha$.

% assumption for \Xi_{S}
On the other hand, the scattering variance ratio $\Xi_{S}$ of the
practical scattering matrix does not have this universality, even
under the condition $A$, $D\gg |B|$, $|C|$ \cite{Zheng2006c}. The
elastic enhancement factor (inverse of $\Xi_S$) is known to be a
function of both the loss parameter $\alpha$ and the coupling, in
general \cite{Savin2006}. Only in the high loss regime
($\alpha\gg1$), can one further assume that the fluctuation part of
the practical impedance $\delta \textbf{Z}$ is much smaller than the
mean part of the practical impedance $\langle \textbf{Z}\rangle$, as
$\delta \textbf{Z} \ll \langle \textbf{Z}\rangle$ and the practical
impedance elements $|Z_{11}|$, $|Z_{22}|\gg |Z_{12}|$, $|Z_{21}|$,
and therefore with Eq.$\ $(1) Zheng \textit{et al.} have derived
\cite{Zheng2006c},
\begin{equation}\label{eq:variance_S_equal}
\Xi_{S} \simeq \Xi_{Z}, \ \ \ \ (\alpha\gg1).
\end{equation}
Note that for the high loss GOE case, $\Xi_{S} \simeq \Xi_{Z} =
1/2$.

\section{Experimental Systems and Results}

\subsection{Three Experimental Systems}

% microwave cavities
In order to experimentally test the predictions above, we use an
Agilent PNA E8364C network analyzer to measure the frequency
dependence of the complex $2\times2$ scattering matrices
$\textbf{S}$ of three two-port microwave scattering enclosures in
the semiclassical limit. To achieve the semiclassical limit, the
typical length scales of the cavities are at least several times
larger than the free-space wavelength making the systems sensitive
to small perturbations. We add perturbing objects (perturbers) in
each wave scattering system and move the perturbers (with the
movement larger or on the scale of the applied wavelength) to create
an ensemble for each wave scattering system. We can convert
$\textbf{S}$ to $\textbf{Z}$ by Eq.$\ $(1), and the characteristic
impedances of the transmission lines connected to the ports are
$Z_{0,11}=Z_{0,22}=50\ [\Omega]$ in all experiments.

% microwave cavities
The first experimental system is a quasi-two-dimensional ray-chaotic
``$1/4$-bowtie-shaped'' microwave billiard illustrated in Fig.$\
$3(a). The cavity is made of copper and has two coupling ports
schematically shown as the red dots in Fig.$\ $3(a). Microwaves are
injected or extracted through each port antenna attached to a
coaxial transmission line, and each antenna is inserted into the
cavity through a small hole (diameter about 0.1 [cm]) in the lid,
similar to previous setups \cite{Yeh2010a, Hemmady2006a,
Hemmady2006c}. Due to the two convex circular arc walls, ray
trajectories are chaotic. This system has previously been used to
test the predictions of RMT \cite{So1995, Gokirmak1998, Chung2000}.
To create an ensemble for statistical analysis, we add two metal
perturbers to the interior of the cavity and randomly move the
perturbers to create 100 different realizations \cite{Yeh2010a,
Yeh2010b}. For each realization, we measure the scattering matrix
over the frequency window ($6 - 18$ [GHz]). The perturbers are
conducting cylinders of diameter 5.1 [cm] and height approximately
equal to that of the cavity (0.7 [cm]).

\begin{figure}

\includegraphics[width=1.6in]{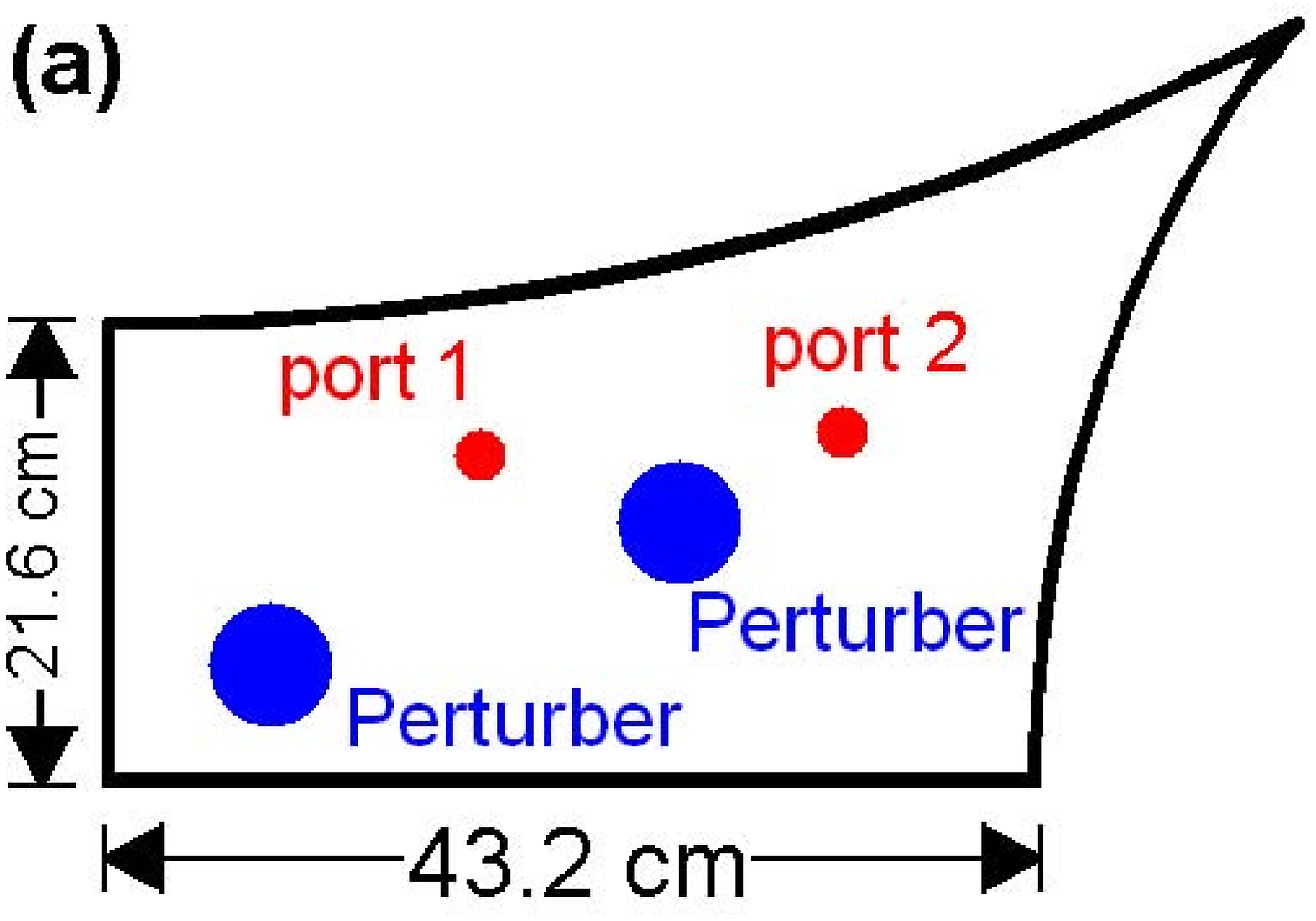}
\includegraphics[width=1.6in]{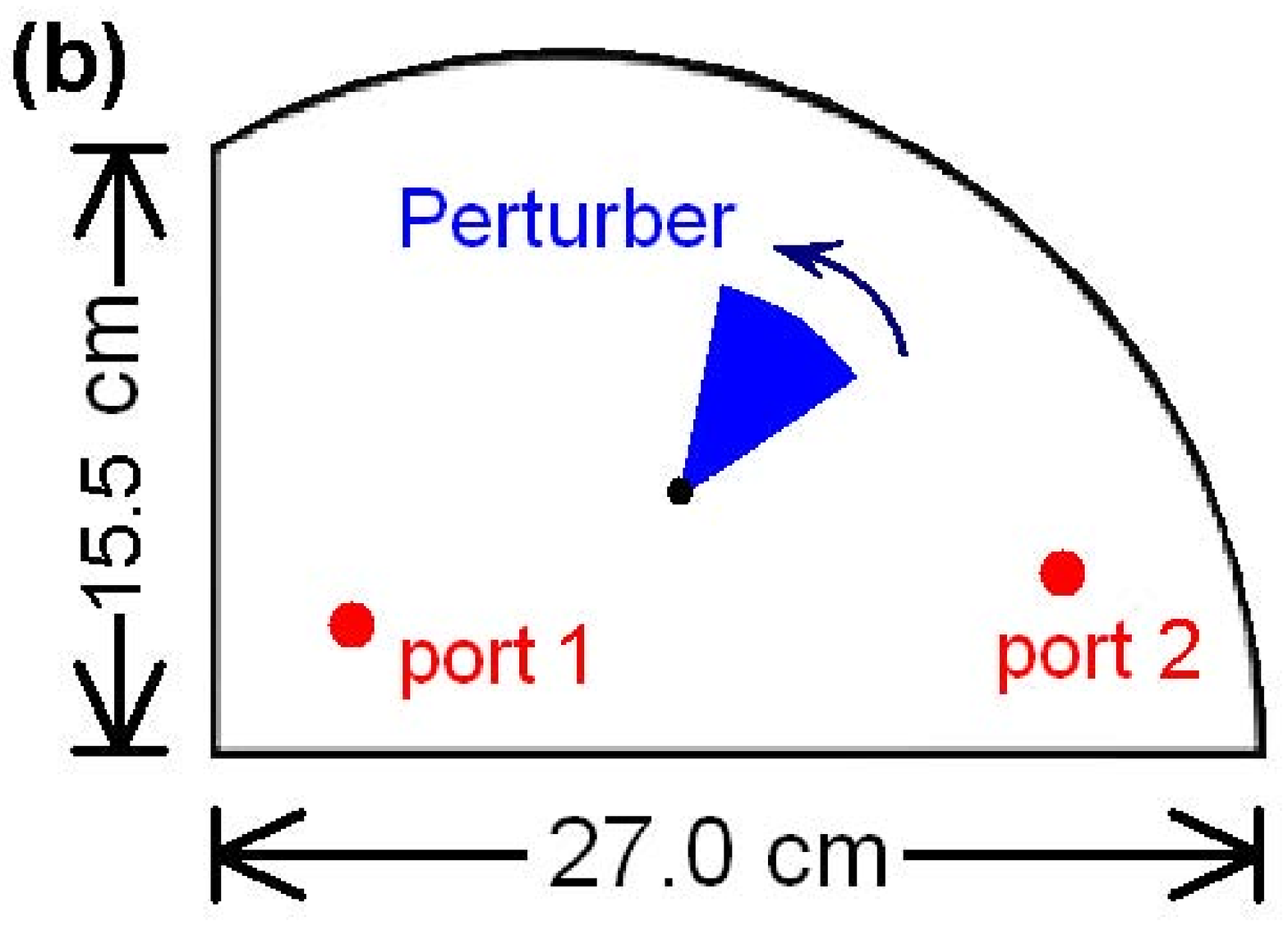}

\caption{(a) The $1/4$-bowtie cavity with the two ports as red dots
and the two metallic perturbers as blue circles. (b) The cut-circle
cavity with the two ports as red dots and the Teflon perturber as
the blue wedge.}

\end{figure}

% superconducting cavity
In order to test the predictions of $\Xi_{Z}$ and $\Xi_{S}$ in the
low loss regime, we have carried out experiments (similar to the
$1/4$-bowtie cavity) in a superconducting microwave cavity,
illustrated in Fig.$\ $3(b). The shape of the cavity is a
symmetry-reduced ``cut-circle'' that shows chaos for ray
trajectories \cite{Yeh2012a, Ree1999, Richter2001, Dietz2006,
Dietz2008}. The superconducting cavity is made of copper with
Pb-plated walls and cooled to a temperature (6.6 [K]) below the
transition temperature of Pb. A Teflon wedge (the blue wedge in
Fig.$\ $3(b)) can be rotated as a ray-splitting perturber inside the
cavity, and we rotate the wedge in $5^{o}$ increments to create an
ensemble of 72 different realizations. Measurements of the
scattering matrix of the superconducting cavity are calibrated by an
\textit{in-situ} broadband cryogenic calibration system (more
experimental details of the cryogenic systems can be found in
\cite{Yeh2013}).

% 3D box and NRL
The previous two wave systems are both quasi-two-dimensional
cavities. We also do experiments in a three-dimensional metal
cavity, which we call the ``GigaBox''
\cite{Taddese2011,Frazier2013}. The GigaBox is approximately a
rectangular microwave resonator with dimensions of length 1.27 [m],
width 1.22 [m], and height 0.65 [m]. The cavity is made of aluminum
and has mode stirrers (a fan formed by aluminum plates) inside it.
The mode stirrers and the irregularities on the surface create a
complicated wave scattering environment. A stepper motor is used to
rotate the mode stirrers to create an ensemble of 199 different
realizations.

\begin{table*}[t]
\centering
\caption{Parameters of the six experimental data sets.}
\begin{tabular}{c|c|c|c|c|c|c}
  \hline
  \hline

  Data Set & I & II & III & IV & V & VI  \\
  \hline

  Cavity            & Cut-circle & Cut-circle & $1/4$-bowtie & $1/4$-bowtie & GigaBox     & GigaBox     \\
  $f_{R}$ [GHz]     & $14 - 16$  & $17 - 19$  & $14 - 16$    & $17 - 19$    & $6.0 - 6.1$ & $9.0 - 9.1$ \\
  $\Delta f$ [MHz]  & 28         & 23         & 10           & 8.6          & 0.031       & 0.014       \\
  $N_{m}$           & 71         & 87         & 200          & 230          & 3200        & 7100        \\
  $N_{r}$           & 72         & 72         & 100          & 100          & 199         & 199         \\
  $\alpha$          & 0.02       & 0.23       & 1.24         & 1.9          & 4.51        & 9.31        \\
  $\Xi_{Z}$         &$0.39\pm0.01$&$0.44\pm0.01$&$0.48\pm0.01$&$0.48\pm0.01$&$0.502\pm0.005$&$0.487\pm0.004$\\
  $\Xi_{Z_{n}}$     &$0.37\pm0.02$&$0.45\pm0.01$&$0.48\pm0.01$&$0.48\pm0.01$&$0.502\pm0.005$&$0.489\pm0.004$\\
  $\Xi_{S}$         &$0.41\pm0.01$&$0.48\pm0.01$&$0.50\pm0.01$&$0.48\pm0.01$&$0.508\pm0.005$&$0.503\pm0.004$\\
  $\Xi_{S_{n}}$     &$0.51\pm0.02$&$0.55\pm0.02$&$0.51\pm0.01$&$0.50\pm0.01$&$0.503\pm0.005$&$0.489\pm0.004$\\

  \end{tabular}
\end{table*}

% data windows
For each of these three microwave systems, we select two frequency
ranges where the condition $A$, $D\gg |B|$, $|C|$ (Eq.$\ $(7)) is
satisfied. The parameters of these six experimental data sets are
shown in Table 1, where $f_{R}$ is the frequency range, $\Delta f$
is the mean frequency spacing of the resonant modes in that range,
$N_{m}$ is the approximate number of modes in the frequency range,
$N_{r}$ is the number of configuration realizations. The first data
set of the cut-circle cavity is measured at temperature 6.6 [K] (the
superconducting case), and the second data set is from the
cut-circle cavity at temperature 270 [K] (the normal metal case).
Note that the GigaBox system has a much higher mode density than the
two quasi-two-dimensional cavities due to its large volume ($V$ =
1.01 [m$^{3}$]), and therefore the smaller frequency range (100
[MHz]) of the GigaBox contains more resonances than the frequency
range (2 [GHz]) of the other two cavities. The loss parameters
$\alpha$ for these data sets are determined as the best-fit
parameter by the method introduced in \cite{Yeh2010b}, which
compares the statistics of the normalized scattering element
$S_{n,12}$ and the prediction of RMT ($S_{rmt,12}$). The averaged
variance ratios ($\Xi_{Z}$, $\Xi_{Z_{n}}$, $\Xi_{S}$, and
$\Xi_{S_{n}}$) and their standard errors of the mean are calculated
from the experimental data, and we introduce the procedures in the
next section.

\subsection{Analysis of the Variance Ratios}

\begin{figure}

\includegraphics[width=3.2in]{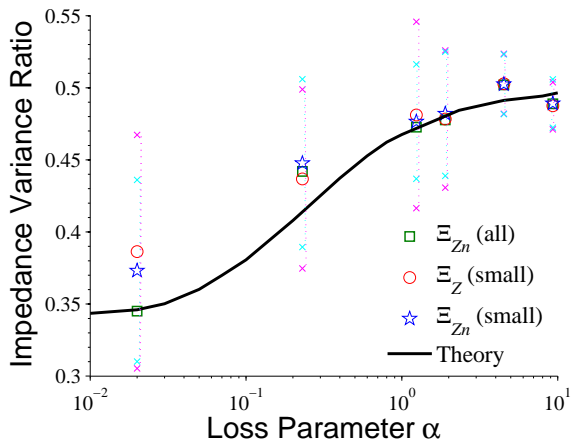}

\caption{The experimental impedance variance ratio versus the loss
parameter $\alpha$. The thick black curve is the analytical formula
$\Xi_{Z_{rmt}}$, Eq.$\ $(5). The green squares are $\Xi_{Z_{n}}$
from the normalized impedance matrix over the whole frequency range.
The red circles are averaged $\Xi_{Z}$, and the pink bars show the
standard deviations of $\Xi_{Z}$ from the practical impedance matrix
over the smaller frequency windows. The blue stars are averaged
$\Xi_{Z_{n}}$, and the light blue bars show the standard deviations
of $\Xi_{Z_{n}}$ from the normalized impedance matrix over the
smaller frequency windows.}

\end{figure}

% exp VRZ
We show the impedance variance ratios of the normalized impedance
matrix $\textbf{Z}_{n}$ and the measured impedance matrix
$\textbf{Z}$ versus the loss parameter $\alpha$ in Fig.$\ $4. As
shown in the finite-size numerical ensembles (Fig.$\ $1), a large
number of samples are critical for accurately determining the
impedance variance ratio, especially in the low loss regime. For
experimental measurement, the number of samples from different
configuration realizations are limited by the remaining correlations
in the experimental data. Therefore, we take the samples for
computing the variance from the ensemble not only with different
configuration realizations but also frequency variations. Note that
in Eqs.$\ $(8) to (11) the variances are taken over the
configuration realizations at a fixed frequency. However, if
$\alpha$ is frequency independent, Maxwell's equations are invariant
to the scaling $f\rightarrow \eta f$ and (length) $\rightarrow
\eta$(length), so that a frequency change can be thought of as
equivalent to a configuration change.

For the normalized impedance matrix $\textbf{Z}_{n}$, the
frequency-dependent nonuniversal features ($A(f)$ and $D(f)$) have
been removed by the RCM, so we can compute the impedance variance
ratio $\Xi_{Z_{n}}$ from variances over the whole frequency range
and all realizations. The results are shown as green squares in
Fig.$\ $4. For the measured impedance matrix $\textbf{Z}$, the
frequency-dependent nonuniversal features remain, so taking
variances over the whole frequency range is not valid. Therefore, we
take a smaller frequency window (1/20 of the whole frequency range
$f_{R}$) instead and assume that the nonuniversal features ($A(f)$
and $D(f)$) are approximately constant in this small frequency
window (100 [MHz] for the cut-circle cavity and the $1/4$-bowtie
cavity, and 5 [MHz] for the GigaBox). With this condition, the
derivation from Eqs.$\ $(8) to (11) is still valid. We compute the
averaged impedance variance ratio $\Xi_{Z}$ of the 20 impedance
variance ratios of the smaller windows and plot the results as red
circles in Fig.$\ $4. For comparison, we also plot the averaged
impedance variance ratio $\Xi_{Z_{n}}$ of small windows for the
normalized impedance matrix as the blue stars. The pink bars (and
the light blue bars) show the standard deviations of the 20 variance
ratios of the measured (and normalized) impedance matrices for the
smaller windows to illustrate the larger fluctuations in the low
loss regime. Note that $1/\sqrt{20}$ of these standard deviations
are the standard errors of the mean shown in the last four rows in
Table 1. Note also that in Fig.$\ $4 the green squares and the blue
stars are both computed from the normalized impedance matrix, and
the only difference is the finite sample size due to the frequency
range for the green squares being 20 times larger than the frequency
range for the blue stars. The values of the blue stars are
systematically larger than the values of the green squares,
especially the lowest loss case. This trend is consistent with the
finite-sample-size deviation illustrated in Fig.$\ $1. Comparing all
three sets of experimental impedance variance ratios, the results in
Fig.$\ $4 agree with the prediction
$\Xi_{Z}=\Xi_{Z_{n}}=\Xi_{Z_{rmt}}$ as a function of the loss
parameter, to the extent permitted by the finite sample sizes.

\begin{figure}

\includegraphics[width=3.2in]{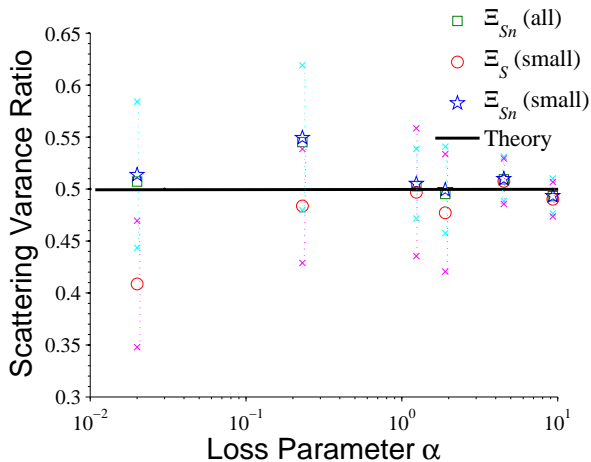}

\caption{The experimental scattering variance ratio versus the loss
parameter $\alpha$. The thick black curve is the theory
$\Xi_{S_{rmt}} = 1/2$. The green squares are $\Xi_{S_{n}}$ from the
normalized scattering matrix over the whole frequency range. The red
circles are averaged $\Xi_{S}$, and the pink bars show the standard
deviations of $\Xi_{S}$ from the practical impedance matrix over the
smaller frequency windows. The blue stars are averaged
$\Xi_{S_{n}}$, and the light blue bars show the standard deviations
of $\Xi_{S_{n}}$ from the normalized impedance matrix over the
smaller frequency windows.}

\end{figure}

% exp VRS
We also convert the impedance matrix to the scattering matrix by
Eq.$\ $(1) and do the same analysis for the scattering variance
ratio. The results are shown in Fig.$\ $5. The experimental results
show that the variance ratios of the normalized scattering matrices
(green squares and blue stars) are consistent with the theoretical
prediction $\Xi_{S_{n}}=\Xi_{S_{rmt}}=1/2$. Note that the measured
scattering variance ratios (red circles and pink bars) tend to be
lower than 1/2, especially in the low loss regime. This trend is
opposite to the finite-sample-size deviation illustrated in Fig.$\
$2 and is due to the nonuniversal features in the wave scattering
system. Zheng \textit{et al.} have shown that the nonuniversal
features (imperfect port coupling) makes the averaged $\Xi_{S} <
1/2$ in the lossless case \cite{Zheng2006c}. Savin \textit{et al.}
have also examined the nonuniversal feature of $\Xi_{S}$ and found
its relationship with the loss parameter in the imperfect coupling
situations \cite{Fyodorov2005, Savin2006}. Hence, the variance
ratios of the scattering matrix $\Xi_{S}$ (red circles) are found
not to be universal, and they depend on the nonuniversal features,
such as the port coupling and short ray trajectories
\cite{Zheng2006c,Savin2006}. Only in the high loss regime
($\alpha\gg1$) is approximately universal behavior of $\Xi_{S}$
observed, such as the two data sets in the GigaBox, where
$\Xi_{S}\simeq1/2$. By comparing Fig.$\ $4 and 5, or the four rows
of variance ratios in Table 1, we see that $\Xi_{S}\simeq\Xi_{Z}$ in
the high loss regime.

% Conclusion & acknowledgment
\section{Conclusion}

In this paper, we analyze the impedance and scattering variance
ratios of complicated wave scattering systems at short wavelength.
Through numerical tests (Fig.$\ $1) and experimental tests in three
microwave systems (Fig.$\ $4), we show that the impedance variance
ratio $\Xi_{Z}$ is a universal function of the loss parameter,
independent of the nonuniversal port coupling and
short-ray-trajectory effects (accounted for in $\textbf{Z}_{avg}$ by
the RCM). On the other hand, the scattering variance ratio $\Xi_{S}$
in general depends on the nonuniversal features (as the low loss
cases in Fig.$\ $5 demonstrate), although it is universal in the
high loss regime.

Comparing with the previous analysis \cite{Zheng2006c}, this work
has two novel contributions. One is that we utilize the
superconducting microwave cavity to test the theoretical predictions
in the low loss regime. The other is that we have utilized the
extended RCM to better account for the nonuniversal features. By
applying the extended RCM to remove the nonuniversal features of the
system, we show that the normalized data ($\Xi_{Z_{n}}$ and
$\Xi_{S_{n}}$) agree with the theoretical predictions
($\Xi_{Z_{rmt}}$ and $\Xi_{S_{rmt}}$) to within the precision
dictated by the finite sample size.

\section*{Acknowledgements}

We thank the group of A. Richter (Technical University of Darmstadt)
for graciously lending us the cut-circle billiard, and H. J. Paik
and M. V. Moody for use of the pulsed tube refrigerator cryostat.
This work is funded by the ONR/Maryland AppEl Center Task A2
(Contract No. N000140911190), the Office of Naval Research (Contract
No. N000141310474), the AFOSR (Grant No. FA95500710049), NSF-GOALI
ECCS-1158644, and the Center for Nanophysics and Advanced Materials
(CNAM).

\end{document}